\documentclass[aps,prb,twocolumn,amsmath,amssymb,superscriptaddress,floatfix]{revtex4-1}

\makeatletter
\def\l@subsubsection#1#2{}
\makeatother

\usepackage{amsfonts}
\usepackage{amsmath}
\usepackage{listings}
\usepackage{enumerate}
\usepackage{latexsym}
\usepackage{bm}
\usepackage{graphicx,float}
\usepackage{siunitx}
\usepackage{subfigure}
\usepackage{braket}
\usepackage{xcolor}
\usepackage{verbatim}
\usepackage{tikz}
\usepackage{physics}
\usepackage{mathtools}

\usepackage{ulem} 
\usepackage{hyperref}
\newcommand{\bs}[1]{\boldsymbol{#1}}

\newcommand{\smbsix}{SmB$_6$}

\newcommand{\bk}{\ensuremath{\bs{k}}}
\newcommand{\twobytwo}[4]{\begin{pmatrix} #1 & #2 \\ #3 & #4\end{pmatrix}}
\newcommand{\EQ}[1]{Eq.~\eqref{#1}}

\definecolor{pyorange}{RGB}{255,127,14}
\definecolor{pyblue}{RGB}{31,119,180}

\usepackage{bbold}
\usepackage{hhline}

\usepackage{titlesec}
\begin{document}
     
\widowpenalty10000
\clubpenalty10000

\title{Topology of SmB$_6$ revisited by means of topological quantum chemistry}

\author{Mikel Iraola}
\email{m.iraola@ifw-dresden.de}
\affiliation{Donostia International Physics Center, 20018 Donostia-San Sebastian, Spain}
\affiliation{Institute for Theoretical Solid State Physics, IFW Dresden, Helmholtzstrasse 20, Dresden, Germany}
\author{Iñigo Robredo}
\affiliation{Donostia International Physics Center, 20018 Donostia-San Sebastian, Spain}
\affiliation{Max Planck Institute for Chemical Physics of Solids, 01187 Dresden, Germany}
\author{Titus Neupert}
\affiliation{Department of Physics, University of Z\"urich, Winterthurerstrasse 190, CH-8057 Z\"urich, Switzerland}
\author{Juan L. Ma\~nes}
\affiliation{Department of Physics, University of the Basque Country UPV/EHU, Apartado 644, 48080 Bilbao, Spain}
\author{Roser Valent\'i}
\email{valenti@itp.uni-frankfurt.de}
\affiliation{Institute for Theoretical Physics, Goethe University Frankfurt, Max-von-Laue-Straße 1, 60438 Frankfurt am Main, Germany}
\author{Maia G. Vergiony}
\email{maia.vergniory@cpfs.mpg.de}
\affiliation{Donostia International Physics Center, 20018 Donostia-San Sebastian, Spain}
\affiliation{Max Planck Institute for Chemical Physics of Solids, 01187 Dresden, Germany}
\date{\today}

\begin{abstract}

The mixed-valence compound SmB$_6$ with  partially filled samarium 4$f$ flat bands 
hybridizing with 5$d$ conduction bands is  a
paramount example of a correlated topological heavy-fermion system. 
In this study we revisit the topology of SmB$_6$ with the band theory paradigm
and uncover previously overlooked aspects resulting from the formation of multiple topological gaps in the electronic structure. By invoking topological quantum chemistry (TQC) we provide  a detailed classification of the strong and crystalline topological features 
that derive from the existence of such topological gaps. To corroborate this classification, we calculate Wilson loops and simulate the surface electronic structure using a minimal tight-binding model, allowing us to describe its surface states and confirm the crystalline topology. We finally discuss  its implications for experiments.

\end{abstract}

\date{\today}

\maketitle

\section{Introduction} \label{Sec: intro}
Due to the presence of correlated electrons and a complex electronic structure,  the number of heavy-fermion materials predicted as topological is still scarce \cite{Chang2017kondo, Li2018, Sundermann2015, Baruselli14, Hagiwara2016, Chang2017kondo}.
Heavy-fermion materials~\cite{Menth1969, Si2013, Steglich1979,Grewe1991,Loehneysen2007} are intermetallic compounds of lanthanides and actinides
with localized $f$ and dispersive $d$ bands near the Fermi surface.
A most discussed type of heavy-fermion material are Kondo insulators~\cite{Menth1969,Si2013}, which undergo a transition into a paramagnetic insulating phase when the temperature is lowered below a critical value.
Importantly, while these materials are strongly-interacting electron systems, their ground states and excitations can be described in terms of highly renormalized f-electrons that hybridize with conduction electrons to form a filled band of quasiparticles \cite{Martin1979, Coleman2015a}.
Alternatively, if the mean occupation of the $f$-orbitals is not close to an integer value, the systems may be classified as mixed valent.
Even though both kinds of systems have been intensively analyzed during the last decades and preliminary research has been done towards a general understanding of their topological properties \cite{Dzero2010,Alexandrov2013,Dzero2012,Dzero2016, Klett20},  there is still a lack of a methodology for the general classification of topological phases in heavy fermion insulators.
This lack of methodology might be one of the reasons why the identification of bulk topological heavy-fermion insulators has not been very successful so far.

In this work, we revisit the topology of \smbsix{} in terms of 
topological quantum chemistry (TQC) \cite{Bradlyn2017, Cano2018, Bradlyn2018} and symmetry-indicators \cite{Po2017, Song2018, Slager2013} via Density Functional Theory (DFT) calculations.
Both, experimental and theoretical studies have labeled this system as a mixed-valence insulator. \cite{Yanase1992, Thunstroem2021, Mizumaki_2009, Cohen1970, Sundermann18, Lee17}
Moreover, experimental analyses have found evidences of surface phenomena which were interpreted as signatures of the presence of topological boundary states. \cite{Yee13, Neupane2013, Xu2013, Jiang2013, Li2020}
These observations are compatible with theoretical analyses which classify \smbsix{} as a strong topological insulator. \cite{Dzero2010, Takimoto2011, Thunstroem2021,Lu2013, Liu2023}
Our analysis has led us to a refined topological classification that considers all crystal symmetries on an equal footing.
We also discuss the origin of topology in terms of the interplay between band representations induced from Sm $4f$ and $5d$-states, and we report the presence of multiple topological gaps close to the Fermi surface.
In addition, we construct a minimal effective tight-binding model able to reproduce the topology of \textit{ab initio} bands.
Based on this model, we simulate the in-gap surface states of the crystal in a slab geometry, and we corroborate the features of crystalline topology.

We note that our approach is based on a single-particle picture of renormalized states, rather than on a multiplet description. \cite{Sundermann18, Thunstroem2021, Denlinger14}
Although these two descriptions are fundamentally different, in the case of SmB$_6$ the symmetry properties of states close to zero energy are identical in both approaches.
As we explain in detail in Appendix~\ref{app: multiplets}, this is due to the fact that the low-energy region of the spectrum of binding energies is dominated by transitions from a singlet to states where the $4f$-shell of Sm is two electrons short from half-filling.
Moreover, previous analyses perfomed via numerical methods beyond DFT regarding the implementation of electron interactions suggest that the single-particle excitation spectrum of \smbsix{} can be described effectively in terms of a quasiparticle picture [see Appendix~\ref{app: interactions} for a more detailed discussion on the effect of interactions].
These facts encouraged us to analyze this material in terms of TQC and symmetry-indicators of topology within the DFT framework, and might inspire the application of this approach to other heavy-fermion materials with similar properties.

We have structured the article in the following way: in Sec.~\ref{sec: formalism} we analyze the general features of the band structures of heavy-fermion insulators from the perspective of TQC, and we describe the way in which the hybridization between $f$ and $d$-bands might lead to topological phases.
Section~\ref{sec: multigap} contains a discussion on the possibility of having multiple cumulative topological bands close to the Fermi level in a range accessible to experimental probes.
In Sec.~\ref{sec: SmB6} we revisit the topological classification of \smbsix{}, and we investigate whether this material is a candidate to exhibit in-gap boundary states close to the Fermi level.
Finally, in Sec.~\ref{sec: conclusions} we present the conclusions and outlook of our work. 
  
\section{Hybridization-driven topology in heavy-fermion insulators} \label{sec: formalism}

A set of orbitals is closed if they span a single-particle Hilbert space invariant under the action of the space group of the crystal.
From a closed set of orbitals one can extract a basis for an infinite-dimensional representation of the space group known as \textit{band representation}. \cite{Zak1980,Zak1981,Zak1982}
The band representation can then be used to describe the transformation properties of bands induced by changing to a basis of Bloch-like combinations of the orbitals.
  An \textit{atomic limit} is associated to a set of bands transforming as a band representation (in reciprocal space).
Furthermore, a band representation that cannot be split into smaller band representations is dubbed an \textit{elementary band representation} (EBR).

According to the formalism of topological quantum chemistry, \cite{Bradlyn2017, Cano2018, Bradlyn2018} if a set of bands does not have an atomic limit, it is topological.
Showing that the representation of a set of bands is not a band representation is then sufficient to demonstrate that they are topological bands.
When this topology is visible to crystal symmetries, it can be inferred from little group irreducible representations (irreps) of bands at maximal \bk-points of the
Brillouin zone (BZ):
if this set of irreps does not coincide with those of any linear combination of EBRs with non-negative integer coefficients, the bands are not related to an atomic limit, and their topology is necessarily nontrivial.

A remarkable feature of mixed valence and (magnetically non-ordered) Kondo insulators containing lanthanide elements is that the low-energy region of their band structure is dominated by the presence of dispersive and heavy bands.
In terms of atomic limits, heavy bands transform as a band representation $\rho_f$ induced from localized $4f$-orbitals of the lanthanide element, while the band representation $\rho_d$ of dispersive bands is induced from spatially extended $5d$-orbitals of the same element.
The hybridization between $4f$ and $5d$-orbitals plays an important role in heavy-fermion insulators since
it is responsible for the fluctuating occupation of $4f$-orbitals. 
When the average occupation of these states is close to an integer value, the system tends to exhibit Kondo behavior, and to become a Kondo insulator below a given critical temperature.
In contrast, if the average occupation corresponds to an intermediate value between two integers, the compound is a mixed-valence material.
The system might, in both cases, have an electronic structure that can be described  in terms of renormalized heavy and dispersive bands, in the absence of magnetic ordering. \cite{Zwicknagl1993}
We set the focus of our discussion on such phases.

\begin{figure}
	\centering
	\includegraphics[width=0.85\linewidth]{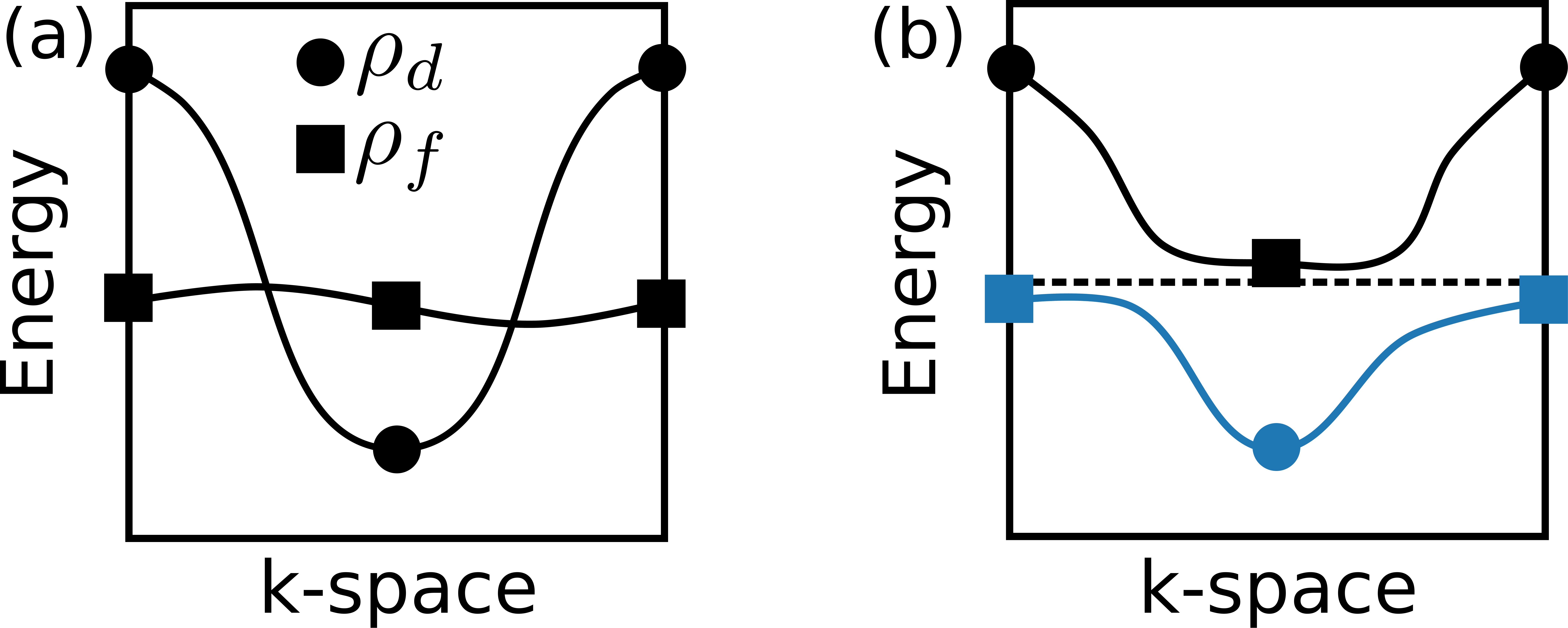}
	\caption{
            Schematic representation of the way the hybridization between $f$ and $d$-bands leads to a heavy-fermion insulator. Circles and squares denote little-group irreps corresponding to $\rho_d$ and $\rho_f$, respectively.
            (a) Band structure if the hybridization were not present.
            (b) Case with hybridization, where a spectral gap separates valence (blue) and conduction (black) bands. Depending on the material, differentiating between irreps of $\rho_f$ and $\rho_d$ might not be possible once the hybridization is considered.
	}	
	\label{fig:cases}
\end{figure}

The hybridization and spin-orbit coupling
(SOC) mediated interplay of $4f$ and $5d$-bands might lead to a nontrivial topology.
Although the hybridization cannot be tuned arbitrarily in a given material, it will be helpful to consider here that we can switch it on and off in order to gain insight into the interplay between dispersive and heavy bands.
We consider as starting point the band structure represented schematically in Fig.~\ref{fig:cases}(a), where the bundle of heavy $4f$-bands intersects the set of dispersive $5d$-bands.
Before considering the hybridization, it is possible to assign every irrep to either $\rho_f$ or $\rho_d$, even when those representations have some irreps in common.
Furthermore, the Fermi level would typically lie on the bundle of intersecting $f$ and $d$-bands, and the system would be a metal.
When the hybridization is turned on, a gap  opens between valence and conduction bands and  the system becomes an insulator, while some  irreps are exchanged  between $\rho_d$ and $\rho_f$.
According to TQC, if the set of irreps of valence bands can no longer be written as a linear combination of EBRs with non-negative integer coefficients, the material becomes  topologically nontrivial.
Identifying this scenario is particularly simple if every irrep  can be related either to $\rho_d$ or $\rho_f$, \textit{i.e.} if these band representations have different irreps.
This is, indeed, the case in \smbsix.

\section{Cumulative topology in heavy-fermion insulators} \label{sec: multigap}

In this section, we discuss the prospect of heavy-fermion insulators to exhibit multiple topological gaps accessible to experimental probes.

\begin{figure}
	\centering
	\includegraphics[width=0.85\linewidth]{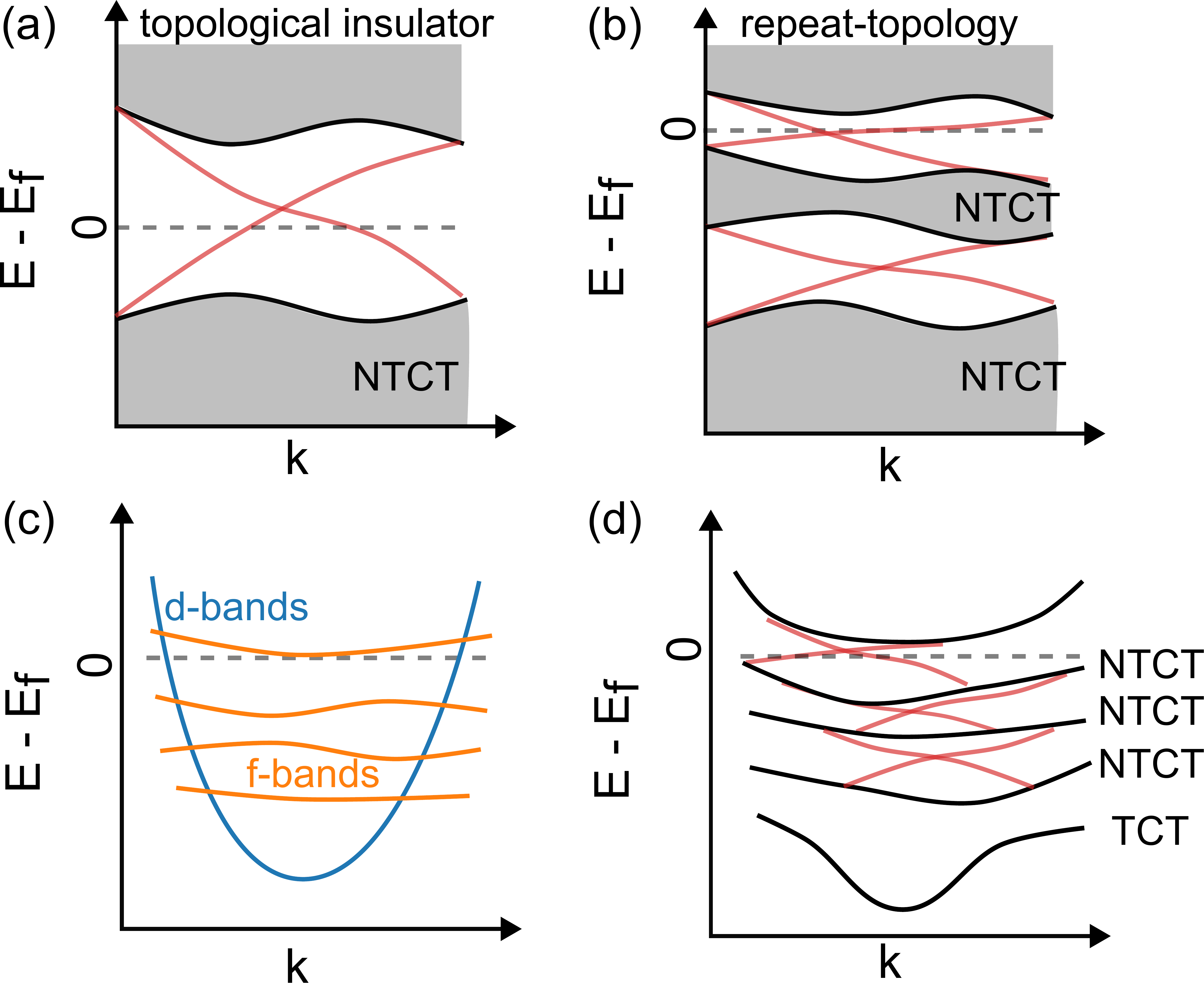}
	\caption{
            Schematic figures representing different cases where boundary in-gap states arise. 
            (a) The case of a generic topological insulator. Boundary states populate the gap at $E_F$ due to the non-trivial cumulative topology (NTCT) of the last set of valence bands.
            (b) Repeat-topological case, where boundary states arise in the gap at $E_F$ and in the next gap below it.
            (c) Bulk band structure of a heavy-fermion system without hybridization. The region around the Fermi level tends to contain a relatively large number of $f$-bands in such systems.
            (d) Boundary band structure of a heavy-fermion system with hybridization. The interplay between $d$ and $f$-bands might give rise to a plethora of bands with NTCT whose boundary projections would be connected by in-gap states located close to the Fermi level.
	}	
	\label{fig:rtopo}
\end{figure}

The cumulative topology of a set of bands is defined as the topology of the group formed by these bands and all lower non-core bands.
The boundary projections of two separated sets of bulk bands are connected by robust in-gap states if the cumulative topology of the lower set of bulk bands is non-trivial. 
In particular, a material shows boundary states connecting valence and conduction bands if the cumulative topology of the last set of valence bands is non-trivial, as it is illustrated in Fig.~\ref{fig:rtopo}(a).

As explained in Ref.~\onlinecite{Vergniory2022}, the presence of in-gap boundary states is not restricted to the separation between conduction and valence bands.
Materials displaying boundary states in the gap at the
Fermi level, and in the first gap below it, are dubbed repeat-topological (RTopo) materials [see Fig.~\ref{fig:rtopo}(b)].
Restricting the definition of repeat-topology in Ref.~\onlinecite{Vergniory2022} to only two gaps is motivated by the fact that the rest of gaps tend to lie at energies that are hardly accessible for experimental probes like angular-resolved photoemission spectroscopy (ARPES).
Nevertheless, materials which exhibit multiple topological gaps could be interesting to explore realizations of multi-gap topology \cite{Wu19, Bastian21, Peng22}.

The number of gaps populated with boundary modes accessible to experimental probes might be especially large in the heavy-fermion phases: in these systems, the number of $f$-bands close to $E_F$ tends to be relatively large, as  represented in Fig.~\ref{fig:rtopo}(c).
The hybridization between $4f$ and $5d$-bands, as well as between $4f$-bands combined with SOC effects, might then yield a considerable number of isolated sets of bands with non-trivial cumulative topology close to $E_F$. According to the discussion above, the boundary-projections of these gaps would be connected by in-gap states [see Fig.~\ref{fig:rtopo}(d)].

\section{SmB$_6$ revisited} \label{sec: SmB6}

\begin{figure}
	\centering
	\includegraphics[width=1.0\linewidth]{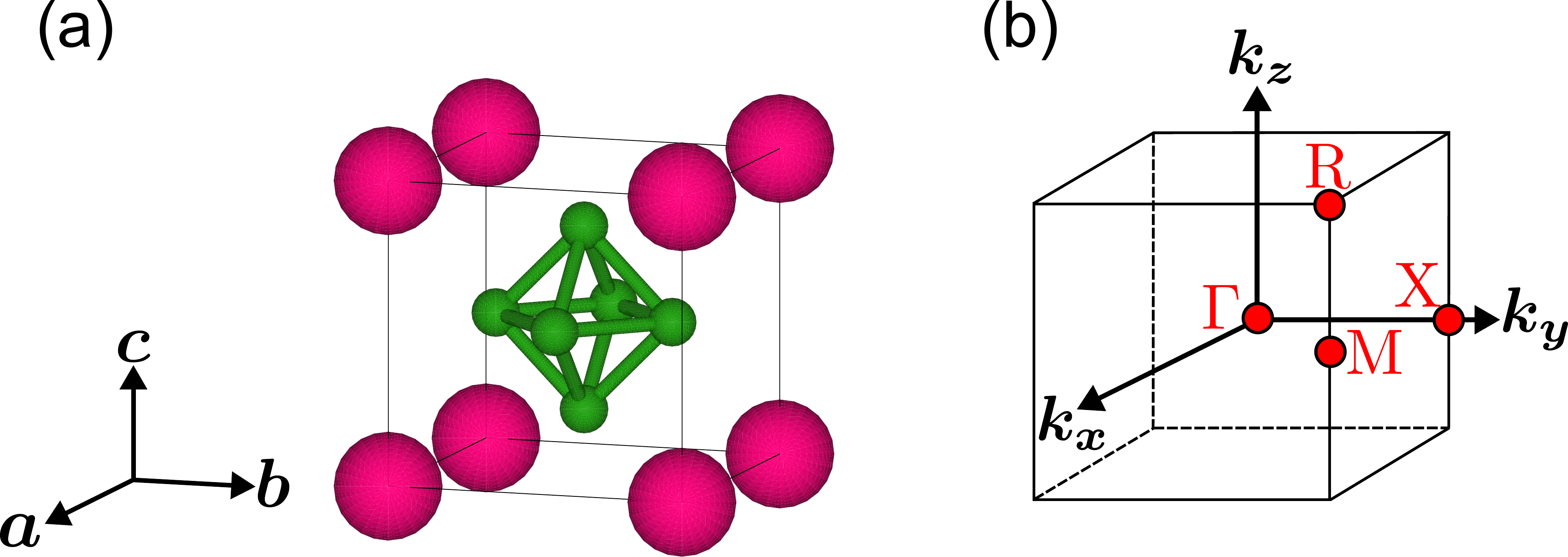}
        \caption{Crystal structure of \smbsix. (a) Primitive unit cell. Sm atoms are shown in magenta (Wyckoff position 1a), while B atoms are shown in green (Wyckoff position 6f). (b) First BZ, where maximal \bk-points are indicated in red.}	
	
	\label{fig: unitcell-BZ}
\end{figure}

\smbsix{} crystallizes in a primitive cubic structure in the space group $Pm\bar{3}m$ (No. 221). 
Our choice of unit cell and BZ are shown in Fig.~\ref{fig: unitcell-BZ}(a) and (b), respectively.
\smbsix{} has been identified as a mixed-valence insulator based on theoretical simulations~\cite{Yanase1992, Thunstroem2021} and experimental evidence~\cite{Mizumaki_2009, Cohen1970, Sundermann18, Lee17}.
Furthermore, theoretical analyses predict \smbsix{} to be a strong-topological insulator. \cite{Dzero2010, Takimoto2011, Thunstroem2021,Lu2013}
This prediction is compatible with the robust surface states reported in experimental probes. \cite{Yee13,Neupane2013,Xu2013,Jiang2013}
Despite evidence suggesting that the compound is topological, this interpretation is still controversial. \cite{Zhu2013, Hlawenka2018}
In this chapter, we revisit the topological classification of \smbsix{} in terms of TQC following the analysis introduced
in the previous section, and we shed light on the origin of its topology.
We support our classification with the analysis of windings in Wilson loop spectra computed with a minimal tight-binding (TB) model that reproduces the topology of the material.
Furthermore, we discuss the potential existence of several cumulative-topological bands close to the Fermi level in \smbsix{}.


\subsection{\textit{Ab initio} band structure and topological classification} \label{subsec: SmB6 ab initio bands}

The ground state electron density and band structure of \smbsix{} have been calculated self-consistently with the Vienna Ab Initio Simulation package \cite{Kresse1996} (VASP).
A plane-wave cutoff of 500 eV 
was used in the self-consistent calculation of the ground-state density and the BZ was sampled with a grid of $9 \times 9 \times 9$. 
Spin-orbit corrections have been included in the calculations. 
The General Gradient Approximation was used for the exchange-correlation term, in the Perdew Burke Ernzerhof \cite{Perdew1996} parametrization.
According to our DFT calculations, the average occupation of the $4f$-shell of Sm in the ground state is $n_f = 5.5$, which is in good agreement with previous works. \cite{Yanase1992, Thunstroem2021, Mizumaki_2009, Cohen1970, Sundermann18, Lee17}
\smbsix{} is thus a mixed-valence insulator where $4f$-states play the role of localized orbitals, whereas $5d$-states act as overlapping orbitals.

\begin{figure}
	\centering
	\includegraphics[width=0.95\linewidth]{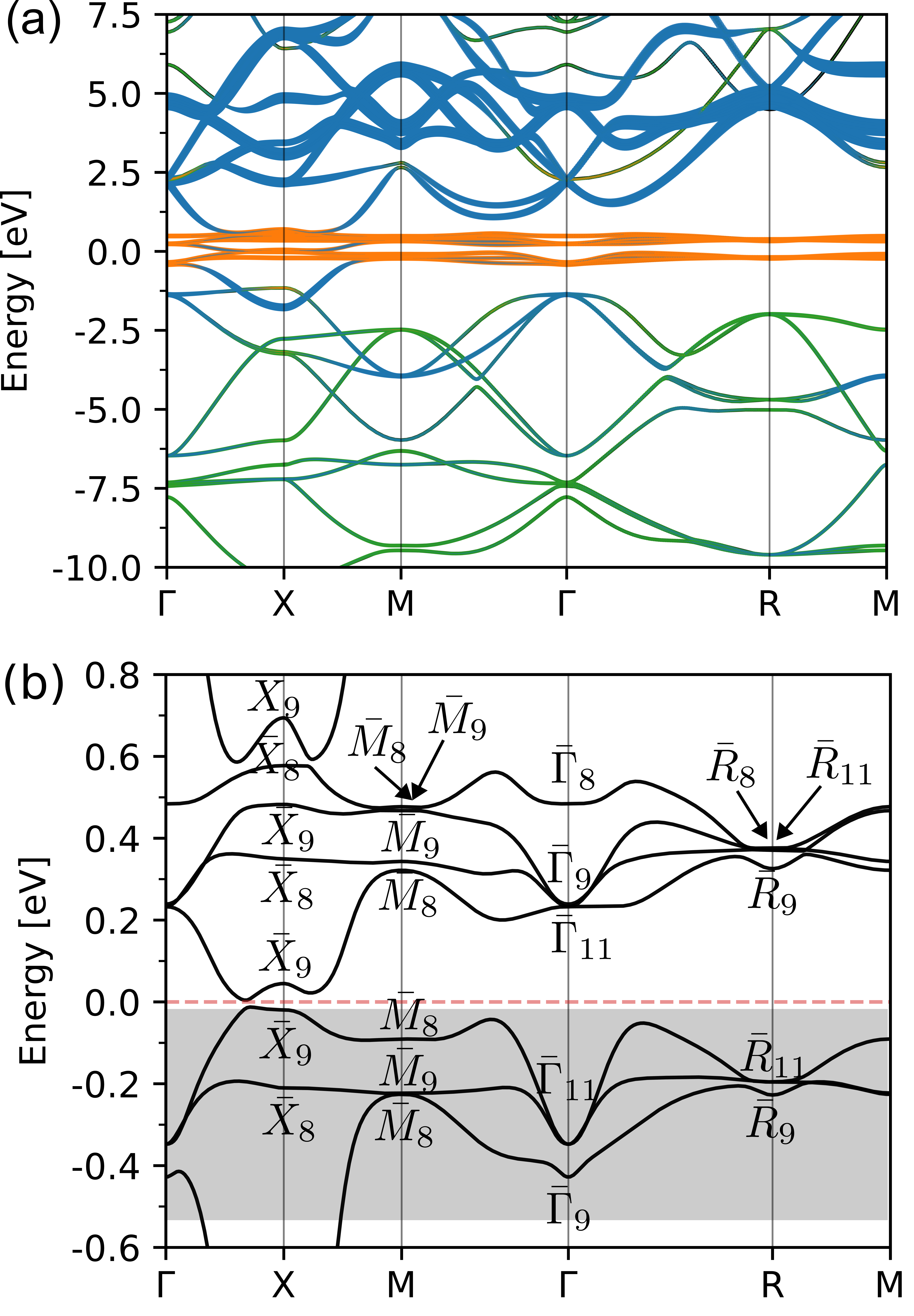}
        \caption{(a) Band structure of \smbsix{} calculated with GGA. Weights of Sm $d$, Sm $f$ and B bands are indicated in blue, orange and green, respectively. (b) $4f$-bands and their little-group irreps. Bands in the grey region correspond to $J=5/2$ states, while those above the Fermi level stem from $J=7/2$ states.}
\label{fig:bandstructure}
\end{figure}

Figure~\ref{fig:bandstructure}(a) shows the band structure of \smbsix{} and weights of Sm $d$ and $f$-states.
Although most $5d$-bands are located above the Fermi level, there is a $5d$-band coming down to $-2$ eV in the line connecting $\Gamma$ to X.
$4f$-orbitals induce heavy (quasi)bands which lie close to the Fermi level and cut through this $5d$-band, thus the low-energy spectrum is dominated by the presence of $5d$ and $4f$-bands, and the interplay between these bands leads to an insulating band structure.
The possibility  of having band crossings at \bk-points not represented in the path shown in this figure is analyzed in Appendix~\ref{app: crossings in SmB6}.


In order to determine if the valence bands are topological, we have calculated their little-group irreps at maximal \bk-points of the BZ with the software \textit{IrRep} \cite{Iraola2022}.
These irreps are shown in Fig.~\ref{fig:bandstructure}(b) for $4f$-bands.
It turns out that the set of irreps of valence bands does not coincide with those of any linear combination of EBRs with positive or zero integer coefficients.
As a result, \smbsix{} is a topological insulator according to the TQC formalism.
Furthermore, we have computed the values for the symmetry-based indicators of topology \cite{Po2017, Song2018} corresponding to valence states via the software \href{https://www.cryst.ehu.es/cgi-bin/cryst/programs/magnetictopo.pl?tipog=gesp}{\textit{CheckTopologicalMat}} \cite{Vergniory2019, Vergniory2022}, which yields the $\mathbb{Z}_4$ indicators $z_4=1$ and $z_{4\pi m} = 3$, the  weak and strong $\mathbb{Z}_2$ indices $(z_{2w,1}, z_{2w,2}, z_{2w,3}; z_2) = (1,1,1;1)$ and the $\mathbb{Z}_8$ index $z_8=5$.
Therefore, the \textit{ab initio} valence bands of \smbsix{} host a strong-topological phase, with features of crystalline topology.
Indeed, the symmetry-indicator $z_{4\pi m} = 3$ implies a mirror-Chern number $C_m|_{k_z=\pi} = 3\ \textrm{mod 4}$ in the $k_z=\pi$ plane, which is compatible with the precise values $C_m|_{k_z=\pi} = 1$ and $3$.
In Sec.~\ref{subsec: TB model SmB6} 
we will confirm that the value for the mirror-Chern number is exactly $C_m|_{k_z=\pi} = 1$.
Our classification is thus able to diagnose in a simple and effective way the features of crystalline topology reported 
early on by Ye \textit{et al.}~\cite{Ye2013}.

This result is consistent with previous theoretical works which classify the phase as strong topological. \cite{Dzero2010, Takimoto2011, Thunstroem2021,Lu2013}
We should emphasize that our classification does not only provide the weak and strong $\mathbb{Z}_2$ topological invariants, \cite{Fu2007} but it also includes additional indices.
Although any odd value for the $z_8$ indicator determines that occupied bands host a strong-topological phase, the interface between two lattices that host phases indicated by different values for $z_8$ might exhibit topological surface states\cite{Song2018}.


\subsection{Low-energy physics and origin of the gap} \label{subsec: cef}

In this section, we follow a group theory and \textit{ab initio} based approach to describe in detail the low-energy part of the band structure of \smbsix{}. 
In particular, we identify the most important couplings that contribute to the existence of the gap between valence and conduction states.


The $f$-shell of the isolated Sm atom consists of 14 orbitals transforming as the representation $D_{3}^- \otimes D_{1/2}^+$ of the symmetry group $O(3)$. Here,  $D_L^p$ denotes the irrep of angular momentum $L$ and parity $p$; for instance, $D_{1/2}^+$ is the spin-representation of $O(3)$.
However, the representation $D_{3}^- \otimes D_{1/2}^+$ is reducible, and can be decomposed in terms of smaller irreducible representations of $O(3)$
\begin{equation}
    D_{3}^- \otimes D_{1/2}^+ (14)  = D_{5/2}^- (6) \oplus D_{7/2}^+ (8),
\end{equation}
where $J=5/2$ and $J=7/2$ are values for the total-angular momentum, and the dimension of each representation is written within brackets.
According to this decompostion, which describes the split produced by SOC from a group theory perspective, the 14-fold degenerate $f$-shell is separated into two groups of 6 and 8 degenerate states.
This separation is visible in the band structure shown in Fig.~\ref{fig:bandstructure}(b), where $J=5/2$ bands are below the Fermi level, while $J=7/2$ states are above it.

Nevertheless, Sm ions are not isolated in \smbsix{}, but they are instead surrounded by B and other Sm ions.
As a consequence, the symmetry group of every Sm site is reduced from $O(3)$ to its site-symmetry group $G_{1a}$, which is isomorphic to the point group $m\bar 3 m$.
$J=5/2$ and $J=7/2$ states transform as the representations of $G_{1a}$ subduced by $D_{5/2}^-$ and $D_{7/2}^-$, which are both reducible and thus decomposable in terms of irreps of the point group
\begin{align}
& D_{5/2}^- \downarrow G_{1a} = \bar E_{2u}(2) \oplus \bar F_u (4), \\
& D_{7/2}^- \downarrow G_{1a} = \bar E_{2u}(2) \oplus \bar E_{1u}(2) \oplus \bar F_u(4).
\end{align}

This decomposition is the group-theory based analysis of the split of the $4f$-shell of Sm due to the surrounding crystal environment, \textit{i.e.} the crystal-field splitting.
According to it, the $J=5/2$ set splits into two groups of 2-fold and 4-fold degenerate states, while $J=7/2$ states separate into a 4-fold degenerate and two 2-fold degenerate sets.
As shown in Fig.~\ref{fig:bandstructure}(b), the splitting produced by the crystal environment is not strong enough to modify significantly the band splittings due to SOC.
This is a consequence of the small off-site overlaps of $4f$-orbitals due to their spatial localization.

Despite the hybridization between $4f$-states being small compared to SOC, it is enough to prevent a crossing between the highest $J=5/2$ and lowest $J=7/2$ bands in the $\Gamma$-X line [see Fig.~\ref{fig:bandstructure}(b)].
Without this hybridization \smbsix{} would be a metal.

The hybridization between Sm $4f$ and $5d$-orbitals is also essential for the gap between valence and conduction states to be finite, as pointed out in Sec.~\ref{subsec: SmB6 ab initio bands}.
In the absence of such hybridization the $5d$-band marked in blue in Fig.~\ref{fig:bandstructure}(a) would not split, and there would be a Fermi surface populated by Bloch states induced from these $5d$-orbitals.

In conclusion, the origin of the spectral gap between valence and conduction states in \smbsix{} is governed by the interplay between the strong SOC of Sm, off-site couplings between $4f$-orbitals, and the hybridization of $4f$ states with Sm $5d$-orbitals.

\subsection{Tight-binding model for \smbsix} \label{subsec: TB model SmB6}

In this section, we present a minimal tight-binding model that reproduces the key topological aspects of the band structure of \smbsix.
The model is based on a simplification of the \textit{ab initio} band structure studied in Sec.~\ref{subsec: SmB6 ab initio bands} which, even if less detailed than the original band structure, provides a clear picture of the surface states of the material and their relation to the underlying topology.


In order to derive the TB model, let us first present the splitting of $5d$-orbitals.
In the isolated Sm ion these orbitals transform as the irrep $D_2^+\otimes D_{1/2}^+$ of the symmetry group $O(3)$.
In \smbsix{} the 10-fold degeneracy of $5d$-orbitals is split  due to its surrounding crystal environment and the strong SOC in Sm.
This splitting is described from a group theoretical perspective as the decomposition of $D_2^+\otimes D_{1/2}^+$ into irreps of the site-symmetry group $G_{1a}$ isomorphic to $m\bar3m$:

\begin{equation}
    D_2^+\otimes D_{1/2}^+(10) = \bar E_{2g}(2) \oplus 2\bar F_g(4).
\end{equation}
Therefore, the $5d$-orbitals separate into a group of 2-fold and two groups of 4-fold degenerate states.
Moreover, the little-group irreps $\bar \Gamma_{10}$ and $\bar X_7$ of the $5d$-band intersected by $4f$-bands coincide with little-group irreps of the band representation $(\bar F_g \uparrow G)_{1a}$.
The rest of $5d$-bands are too far  from the Fermi level to play any role in the topology.
This motivates us to restrict the set of $5d$-orbitals included in the TB model to the set of four orbitals transforming as the irrep $\bar F_g$ of the site-symmetry group $G_{1a}$.

To make an efficient choice of $4f$-states, we first note that the contribution of the $5d$-bands to the valence states at high symmetry points (HSPs) is limited to the $X$-point.
Moreover, if there were no $5d$-bands close to the Fermi level, the set of valence  irreps at X would be $\{\bar X_8, \bar 2X_9\}$, with  $\bar X_9$ the irrep of the last valence band.
However, due to the interplay between $5d$ and $4f$-bands, the set of valence irreps is $\{\bar X_7, \bar X_8, \bar X_9\}$ instead, with $\bar X_7$ coming from the $5d$ bands through the mechanism visualized in Fig.~\ref{fig:cases}. 
Thus, effectively a band inversion has taken place at X such that the irreps $\bar X_7$ and $\bar X_9$ become part of valence and conduction states respectively.
This observation suggests that the topological phase could be reproduced by considering only the set of $4f$-bands connected to the irrep $\bar X_9$.
Those bands have the irreps $\bar \Gamma_{11}$ at $\Gamma$, and the pair $\{\bar X_8, \bar X_9\}$ at X, and their irreps coincide with those of the band representation $(\bar F_u \uparrow G)_{1a}$.
The rest of $4f$-bands do not play an essential role in  the effective band inversion and can be safely left out of the TB model.
Therefore, we restrict the set of $4f$-orbitals included in the TB model to those transforming as the irrep $\bar F_u$ of the site-symmetry group $G_{1a}$.

Altogether, we consider eight spinful Wannier functions sitting at WP 1a.
Four of them transform as the irrep $\bar F_g$ under the action of the site-symmetry group $G_{1a} = m\bar{3}m$, while the rest transform as the irrep $\bar F_u$.
The induced  EBRs $(\bar F_g \uparrow G)_{1a}$ and $(\bar F_u \uparrow G)_{1a}$  contain the following little-group irreps at maximal \bk-points
\begin{eqnarray}
    & (\bar F_g \uparrow G)_{1a} : \{\bar \Gamma_{10}, \bar X_{6} \oplus \bar X_{7}, \bar M_6 \oplus \bar M_7, \bar R_{10}\} , \\
    & (\bar F_u \uparrow G)_{1a} : \{\bar \Gamma_{11}, \bar X_{8} \oplus \bar X_{9}, \bar M_8 \oplus \bar M_9, \bar R_{11}\}.
\end{eqnarray}

In order to deal efficiently with the constrains set by symmetries on the parameters of the model, and to write down the Hamiltonian, it is convenient to consider the decomposition of these spinful representations as the product of the spin representation $S=\bar E_{1g}$ and a spinless representation
\begin{eqnarray}
	& \bar{F}_g = E_g \otimes S,
	\label{eq: F_g decomposition spin rep} \\
	& \bar{F}_u = E_u \otimes S.
	\label{eq: F_u decomposition spin rep}
\end{eqnarray}

Based on these decompositions, the tight-binding Hamiltonian\cite{note_hamiltonian} in reciprocal space can be written in the following way

\begin{widetext}
\small
\begin{equation} \label{eq: TB Hamiltonian}
\begin{split}
H&(\bk) = \\
& [\epsilon_d + t_1\sum_{i=x,y,z}\cos k_i] (\nu_0+\nu_3)\otimes\tau_0\otimes\sigma_0 \\
& + t_2 [ \cos(k_x+k_y)+\cos(k_x-k_y)+\cos(k_y+k_z)+\cos(k_y-k_z)+\cos(k_y+k_x)+\cos(k_y-k_x) ](\nu_0+\nu_3)\otimes\tau_0\otimes\sigma_0\\
& + t_3 \{ \cos(k_x+k_y)+\cos(k_x-k_y)+e^{i2\phi}[\cos(k_y+k_z)+\cos(k_y-k_z)]+e^{-i2\phi}[\cos(k_y+k_x)+\cos(k_y-k_x)]\}(\nu_0+\nu_3)\otimes\tau_1\otimes\sigma_0\\
& + t_4 \{ \cos(k_x+k_y)+\cos(k_x-k_y)+e^{i\phi}[\cos(k_y+k_z)+\cos(k_y-k_z)]+e^{-i\phi}[\cos(k_y+k_x)+\cos(k_y-k_x)]\}(\nu_0-\nu_3)\otimes\tau_1\otimes\sigma_0\\
& + V \sum_{i}\sin k_i\ \nu_1\otimes\tau_0\otimes\sigma_i.
\end{split}
\end{equation}
\normalsize
\end{widetext}
Here $\nu_i$ are the Pauli matrices for the $d$ and $f$ sublattice degree of freedom, while $\tau_i$ are the Pauli matrices for the states in the basis of irreps $E_g$ and $E_u$ in Eqs.~\eqref{eq: F_g decomposition spin rep} and~ \eqref{eq: F_u decomposition spin rep},   $\sigma_i$ are the Pauli matrices for spin and 
$\phi=4\pi/3$. 
The term proportional to $\epsilon_d$ is the on-site energy of $d$-orbitals, while the origin of energies is chosen so that 
$\epsilon_f=0$. The second line in Eq.~\eqref{eq: TB Hamiltonian} represents nearest-neighbor hoppings between $d$-orbitals, the third and fourth lines define next-nearest neighbor (NNN) hoppings between this kind of orbitals, and the fifth line accounts for NNN couplings between $f$-states. 
Lastly, the term proportional to $V$ is responsible for the hybridization between $d$ and $f$-orbitals.
See Appendix \ref{app: construction TB} and Fig.~\ref{fig:SmB6 TB bands} for the details of the construction and band structure of the model.

In order to corroborate the mirror-Chern number $C_m|_{k_z=\pi}=1$ mod.\ 4 predicted from the values for symmetry-indicators, we have analyzed the Wilson loop operator ${W(k_y, k_z=\pi)}$ defined on the $k_z=\pi$ plane as \cite{Bradlyn2022}

\begin{equation}
    \begin{split}
    W(k_y,k_z=\pi) &= \prod_{\delta: 0\rightarrow 2\pi} P_\pi(2\pi-\delta,k_y) \\
                   &= P_\pi(2\pi,k_y)P_\pi(2\pi-\delta,k_y)\dots P_\pi(0,k_y),
    \end{split}
\end{equation}
where $\delta$ is a number varying in infinitesimally small steps, and $P_\pi(k_x,k_y)=\sum_n \ketbra{\psi_n(k_x,k_y,\pi)}{\psi_n(k_x,k_y,\pi)}$ is the projector onto valence states at a point $\bk=(k_x,k_y,\pi)$ of the $k_z=\pi$ plane.
The little group of every \bk-point in this plane contains the mirror reflection $M_z$, whose action on ${W(k_y, k_z=\pi)}$ is given by:

\begin{equation} \label{eq: mirror acting on WL}
    M_z W(k_y,k_z=\pi) M_z^{-1} = \prod_{\delta: 0\rightarrow 2\pi} M_z P_\pi(2\pi-\delta, k_y)M_z^{-1},
\end{equation}
where we have inserted the identity $M_z^{-1}M_z=\mathbb{1}$ between every pair of projectors.
From the periodicity of Bloch states in reciprocal states it follows that $M_z$ commutes with every projector, $M_z P_\pi(k_x,k_y) M_z^{-1} = P_{-\pi}(k_x,k_y) = P_\pi(k_x,k_y)$. Therefore, Eq.~\eqref{eq: mirror acting on WL} reduces to the commutation relation

\begin{equation}
    M_z W(k_y,k_z=\pi) M_z^{-1} = W(k_y,k_z=\pi).
\end{equation}
As a consequence of this relation, it is possible to find a basis of states which are simultaneously eigenstates of $M_z$ and $W(k_y,k_z=\pi)$, and hence the eigenstates of the Wilson loop operator can be separated by their $M_z$-eigenvalue.

\begin{figure} 	
    \centering
	\includegraphics[width=1.0\linewidth]{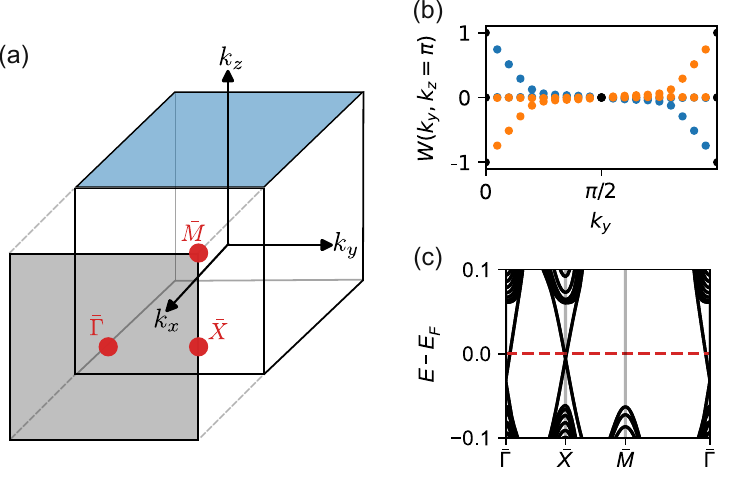}
	\caption{
		(a) BZ of the crystal with the plane $k_z=\pi$ -- where the mirror-Chern number $C_m|_{k_z=\pi}$ is defined -- indicated in blue. The grey plane denotes the BZ of the slab where surface states are computed.
		(b) Spectrum of the Wilson loop $W(k_y, k_z=\pi)$. In orange (blue), eigenstates corresponding to $M_z$ eigenvalue +1 (-1).
		(c) Spectrum of the (100)-surface corresponding to a slab that is finite along $x$-direction with $N=20$ unit cells. The red-dashed line indicates the Fermi level of the bulk band structure.
	}\label{fig:WL_surface-states}

\end{figure}

Figure~\ref{fig:WL_surface-states}(b) 
shows the spectrum of $W(k_y,k_z=\pi)$ calculated with the package \textit{PythTB} \cite{coh_python_2013}.
For each eigenvalue of $M_z$, there is a curve in the spectrum of the Wilson loop which winds once as we go through the BZ.
Based on the fact that the number of windings in Wilson loop spectra coincides with the mirror-Chern number in this family of topological phases, \cite{Bradlyn2022, Neupert2018} 
the number of windings found here corroborates the mirror-Chern number $C_m|_{k_z=\pi} = 1$~mod~4 predicted from the symmetry-indicators of \textit{ab initio} valence bands.

Figure~\ref{fig:WL_surface-states}(c) shows the bands calculated for a finite (along the $x$-axis) slab.
The resulting spectrum is consistent with the strong topological nature predicted from \textit{ab initio} calculations, as the number of surface Dirac cones at the time-reversal invariant momenta of the projected Brillouin zone is odd. \cite{Hasan2010, Fu2007a} 
Additionally, the spectrum exhibits a single Dirac cone along the path connecting $\bar X$ and $\bar M$ points, which agrees with the anticipated implications \cite{Neupert2018, Teo2008, Hsieh2012} of possessing a mirror-Chern number ${C_m|_{k_z=\pi} = 1}$ mod.\ 4. 
The obtained surface spectrum is consistent with experimental observations \cite{Neupane2013, Xu2013, Jiang2013} and previous theoretical simulations of surface states \cite{Lu2013, Thunstroem2021, Takimoto2011,Alexandrov2013, Ye2013}.

\subsection{Multigap topology in \smbsix} \label{subsec: multigap SmB6}

In this section, we comment on the potential of \smbsix{} to host multiple topological gaps close to the Fermi level.
Since heavy bands tend to be confined in a narrow energy window, gaps separating them are usually small.
Hence, their presence in an \textit{ab initio} band structure might depend on the details of the calculation, like the approximation used for the exchange correlation term.
For instance, the GGA band structure in Fig.~\ref{fig:bandstructure} does not display clear multiple topological gaps close to the Fermi level, while the band structure
obtained using the modified-Becke-Johnson (MBJ) parametrization  \cite{Becke06} in Fig.~\ref{fig:mbj multigap} does (see Appendix~\ref{app: interactions} for details).

Both, the interplay between $4f$ and $5d$-bands, and the topological classification of valence bands, are analogous in the electronic structures computed with GGA and MBJ.
Moreover, the MBJ spectrum shows close to the Fermi level three gaps separating bands with non-trivial cumulative topology, whose surface projections are expected to host in-gap states.
The classification of the topology of these gaps is shown in Tab.~\ref{tab:gaps}.

The gap indicated in red separates valence and conduction bands, dictating that valence bands are topological.
According to the discussion in Sec.~\ref{sec: multigap}, since the first gap below the Fermi level (indicated in blue) is also topological, \smbsix{} is a repeat-topological material within the MBJ approximation.
In addition, the first set of conduction bands also display non-trivial cummulative topology, thus we could also expect the first gap above the Fermi level to  exhibit in-gap states. Moreover, the fact that this gap is indirect -- i.e. the maximum of the lower band is smaller than the minimum of the band on top -- makes it a promising testbed for an experimental confirmation of the validity of the single-particle description of \smbsix.

The topological gaps shown in Fig.~\ref{fig:mbj multigap} share the same values for symmetry-indicators due to the fact that the irreps of heavy bands involved in the displayed energy range coincide with the irreps of atomic limits coming from $4f$-orbitals.
Generally, the interplay between $4f$-bands might lead to heavy bands with non-zero values for the symmetry indicators, which could yield gaps with different topology.
These results suggest that \smbsix{} might be a promising material to investigate the presence of topological gaps close to the Fermi level.


\begin{figure}
        \centering
	\includegraphics[width=0.9\linewidth]{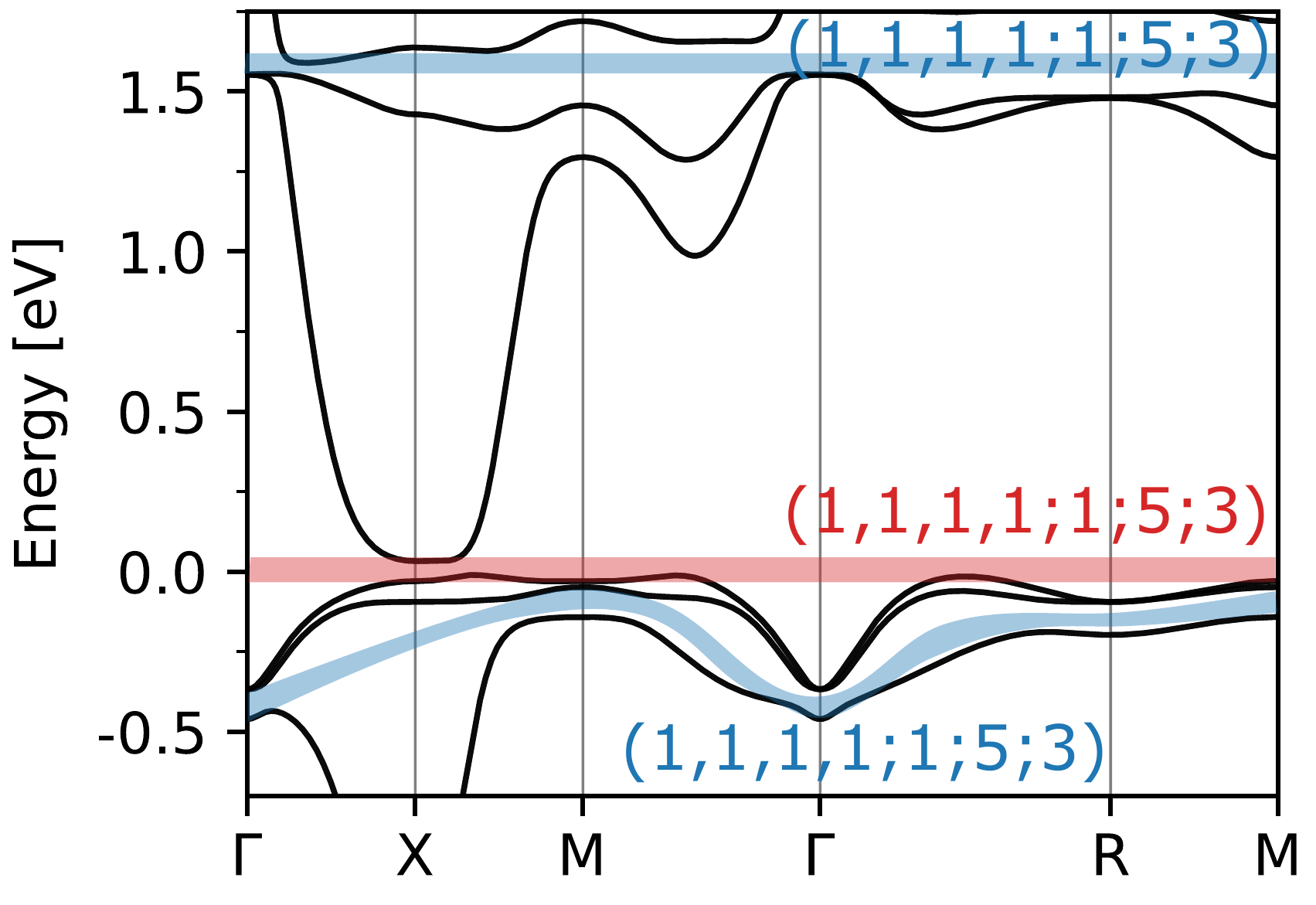}
	\caption{
            Band structure of \smbsix{} calculated with MBJ functionals. Gaps separating bands with non-trivial cumulative topology from bands above them are indicated in colors. The gap in red separates valence and conduction bands, while blue gaps separate two sets of conduction or valence bands. Cumulative values for the symmetry-indicators of topology corresponding to bands below each gap are indicated as $(z_{2w,1}, z_{2w,2}, z_{2w,3}, z_{2}; z_4; z_{4\pi m}; z_8)$.
	}
        \label{fig:mbj multigap}
\end{figure}

Different factors might influence the ease to observe in-gap states related to these topological gaps.
First, their form and presence on a particular boundary of the crystal might depend on the microscopic details of the actual boundary. In the case of \smbsix{} these could be the presence of Sm$_2$O$_3$ impurities, the tendency of the B terminated surface to attract -- due to electrostatic interaction -- Sm atoms forming patterns of altered periodicity, or the difficulty to obtain flat surfaces via cleavage \cite{Rosa20, Wirth21,Ohtsubo2019, Luo21, Ohtsubo2022}.
Second, whether in-gap states are isolated from bulk bands might also be important for the possibility of observing these boundary modes.
In fact, if the  minimum energy of the first band above the gap is smaller than the maximum of the first band below it, 
the projections of bulk bands on a surface might ovelap and prevent the existence of a spectral gap populated exclusively by boundary in-gap states.
Furthermore, some heavy-fermion insulators might show a breakdown of the Kondo coherence \cite{Alexandrov2013, Alexandrov2015} or present a difference in the valence between the bulk and surface, which might make unclear the manifestation of the bulk-boundary correspondence in these systems.

Although we focused on gaps arising from the interplay between dispersive $5d$ and $4f$ bands -- since this is an aspect particular to heavy-fermion systems -- the band structure of the material might contain additional topological gaps of different origin.
Indeed, the band structure of \smbsix{} exhibits topological bands originating from the interplay between boron bands, as we discuss in Appendix~\ref{app: valence topological gaps}.

\section{Discussion and Conclusions} \label{sec: conclusions}

In this work we have revisited the topology of \smbsix{} within the framework of TQC and symmetry-indicators by performing \textit{ab initio} DFT calculations.
While state-of-the-art many-body methods beyond DFT provide an accurate description
of the electronic properties of \smbsix{},~\cite{Thunstroem2021, Lu2013} our DFT calculations also reproduce the mixed-valence behavior of this material, and yield valence bands whose topology is consistent with previous analyses  and experimental observations.  
We have reached a detailed classification of \smbsix{} as strong-topological insulator with features of crystalline topology, and constructed a minimal tight-binding model which reproduces the topology of \textit{ab initio} valence bands.
We have further used this model to corroborate the mirror-Chern number $C_m|_{k_z=\pi}=1\ \textrm{mod 4}$ predicted from the symmetry-indicators, and to simulate the surface band structure of the crystal. Our simulations are consistend with ARPES experiments \cite{Neupane2013, Xu2013,Jiang2013,Zhang2012}.

Regarding the properties of crystalline topology of \smbsix{}, although we focused here  on the mirror-Chern number that follows directly from the $z_{4\pi m}$ symmetry-indicator, the presence of additional topological invariants with non-trivial values cannot be discarded.
In fact, the surface states in Fig.~\ref{fig:WL_surface-states}(c)
would be consistent with $C_m|_{k_z=0}=2$~mod~4 mirror-Chern number, and with the crystalline features studied in Refs.~\cite{Ye2013,Zhang2012}.

Moreover, we have suggested heavy-fermion insulators as potential candidates to host multiple topological in-gap states close to the Fermi level.
Based on this suggestion, and the success of our approach in leading to a detailed classification of the topology of valence bands, we hope that the current work might motivate an intesive search of topological phases in heavy-fermion materials.
Such a renewed interest could induce the discovery of novel topological materials, as well as help diagnosing as topological some phases that had been previously classified as trivial.

We suggest that the inclusion of magnetic elementary band correpresentations\cite{Elcoro2021} could make the TQC based approach applicable to magnetic topological heavy-fermion materials, as long as their electronic structure can be described in terms of renormalized bands.
It remains unexplored to which extent our formalism could be applicable to interacting heavy-fermion phases that are not adiabatically connected to band-insulators \cite{Iraola2021, Lessnich2021}.

\section{Acknowledgements}

M.I. thanks S. Wirth  and K. Held for enlightening discussions. The work of M.I. was funded by the European Union NextGenerationEU/PRTR-C17.I1, as well as by the IKUR Strategy under the collaboration agreement between Ikerbasque Foundation and DIPC on behalf of the Department of Education of the Basque Government. M.G.V. thanks partial support from the Ministry of Economic Affairs and Digital Transformation of the Spanish Government through the QUANTUM ENIA project call – Quantum Spain project, and by the European Union through the Recovery, Transformation and Resilience Plan – NextGenerationEU within the framework of the Digital Spain 2026 Agenda and European Research Council (ERC) grant agreement no. 10102083. M.G.V. and M.I. thanks support to the Spanish Ministerio de Ciencia e Innovacion (grant PID2022-142008NB-I00). M.G.V. R.V. and T.N. acknowledge support of the Deutsche Forschungsgemeinschaft (DFG, German Research Foundation) GA 3314/1-1 -FOR 5249 (QUAST). M.G.V and R.V. are also grateful to the National Science Foundation under Grants No. NSF PHY-1748958 and PHY-2309135.

\appendix

\section{Comparison between classifications based on single-particle and multiparticle states} \label{app: multiplets}

In this section we will argue that symmetry properties of multiplet and single-particle states close to zero energy are identical in crystals with a shell two electrons short of being half filled.
We will follow an approach based on Hund's rules.

\begin{figure}
	\centering
	\includegraphics[width=0.85\linewidth]{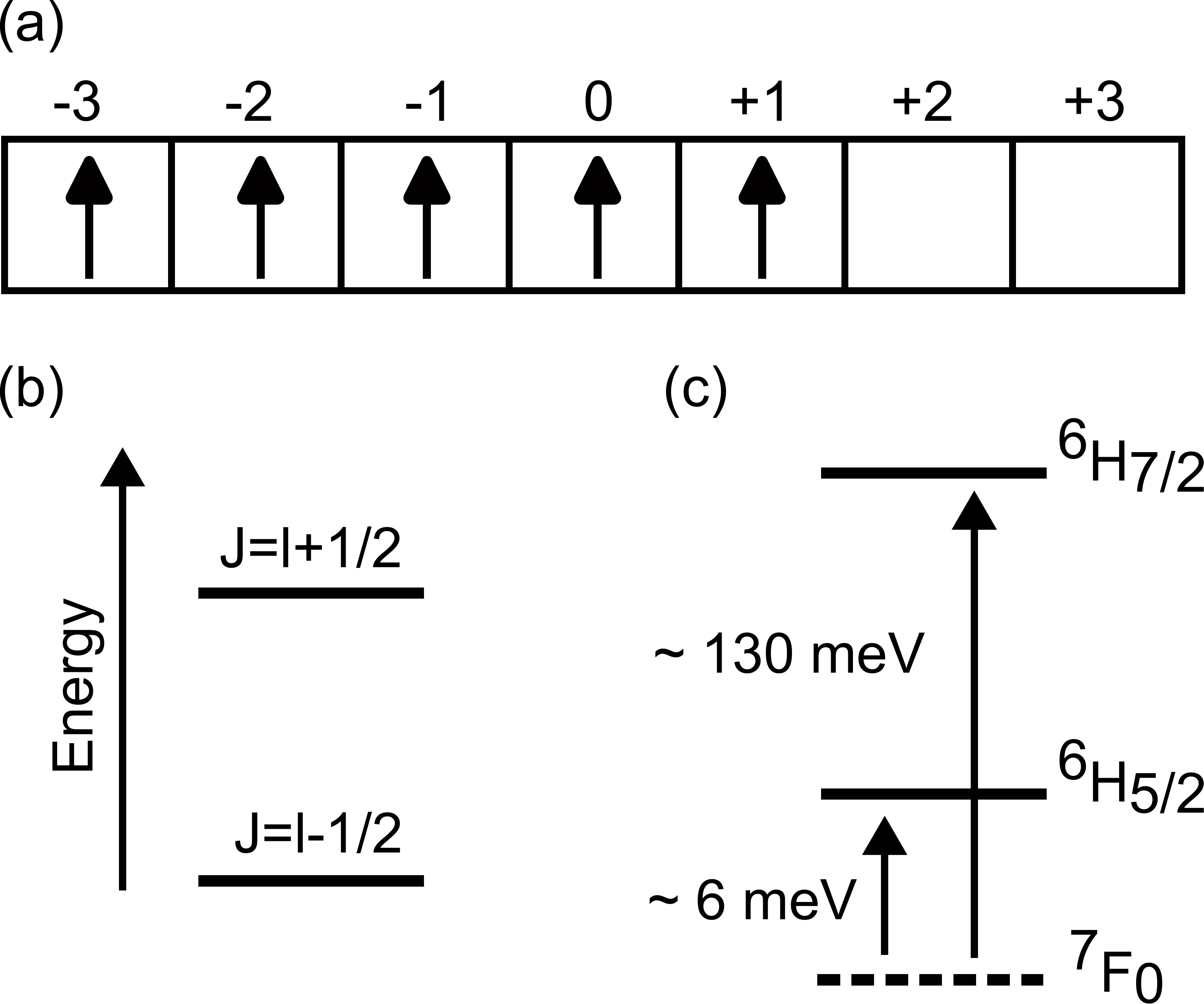}
	\caption{
            (a) Diagram for the application of Hund's rules to predict the ground state of a (f-)shell two electrons short from half-filling.
            (b) Total angular momentum's quantum numbers predicted by Hund's rules for the ground state and first excited state.
            (c) The case of \smbsix. Solid lines indicate the ground state and first excited multiplet states for an occupation of the $f$-shell two electrons short from half-filling, while the dashed line is the singlet ground-state multiplet of the shell containing six electrons. Binding energies for the transition from the singlet to the multiplets of six electrons are represented \cite{Sundermann18, Thunstroem2021, Denlinger14, carnall68, Liu2023}.
	}	
	\label{fig:multiplets}
\end{figure}

Let us consider an atom whose shell of orbital-angular momentum number $l$ is two electrons short from half-filling, \textit{i.e} it contains $2l-1$ electrons, as illustrated in Fig.~\ref{fig:multiplets}(a).
According to Hund's first rule, the multiplet with lowest energy is the one with largest value for the quantum number $S$ of the total spin, which is $S=l-1/2$ for the case considered here.
On the other hand, according to Hund's second rule the multiplet of lowest energy also has the largest allowed value for the total orbital-momentum number $L$ consistent with Hund's first rule, which is $L=2l-1$.
This combination of spin and orbital momenta leads to a total momentum whose quantum number $J$ could take the values $J=l-1/2, l+1/2, \cdots, 3(l-1/2)$.
Hund's third rule states that, for shells with an occupation smaller than half-filling, the multiplet of lowest energy has in correspondence the minimum value of $J$.
Therefore, the term of smallest energy is $J=l-1/2$.
Furthermore, we could expect the first excited state to be $J=l+1/2$, as it is shown in the diagram of Fig.~\ref{fig:multiplets}(b).

Regarding single-particle states stemming from the shell of orbital momentum number $l$, their total angular momentum's quantum number $j$ can take the values $j=l\pm1/2$.
These numbers coincide with those of the two multiplets of lowest energy, and therefore single-particle and multiplet states transform identically under symmetries.

Although this property holds for atoms with a shell two electrons short from half-filling, it is also valid for mixed-valence systems whose spectrum of binding energies is dominated, in the low energy region, by transitions from a singlet to multiplets with this filling.
This is indeed the case of \smbsix{}, where the the smallest binding energies correspond to transitions from the singlet ground state $^7F_0$ of $n=6$ electrons to the multiplet states $^6H_{5/2}$ and $^6H_{7/2}$ of $n=5$ electrons \cite{Sundermann18, Thunstroem2021, Denlinger14} [see Fig.~\ref{fig:multiplets}(c)].
At the same time, single-particle states around the Fermi level are originated from $4f$-orbitals with $l=3$, thus they are states of $j=5/2$ and $j=7/2$ total angular momentum.
Multiplets and single-particle states close to zero energy share, therefore, total angular momentum numbers.

The validity of this result relies on the applicability of Hund's rules.
We could expect these rules to be valid for lanthanide and actinide elements in mixed-valence and Kondo insulators, as their $f$-shells are well localized deep inside the ions and can not be drastically affected by the crystal environment.
Indeed, this is the case for \smbsix{} as suggested by the fact that the binding energies calculated in Refs.\cite{Sundermann18, Thunstroem2021, Denlinger14} for the crystal system are close to those  computed in Ref.\cite{carnall68} for isolated Sm$^{3+}$ ions.

\section{Effect of strong interactions} \label{app: interactions}

Since \smbsix{} is a material where interactions are predicted to be strong, we also calculated the band structure with the Modified-Becke-Johnson (MBJ) parametrization for the exchange-correlation functional \cite{Becke06}.
Although heavy-bands occupy a broader range of energy in MBJ band structure, the general features discussed in the main text are identical for both parametrizations.
In particular, the interplay between Sm $5d$ and $4f$ bands, and topological classification of valence bands is similar with both parametrizations.
We should mention that the bulk band spectrum obtained with MBJ exhibits more separated subsets of bands with non-trivial cumulative topology than the GGA spectrum.

\begin{figure}
	\centering
	\includegraphics[width=1.0\linewidth]{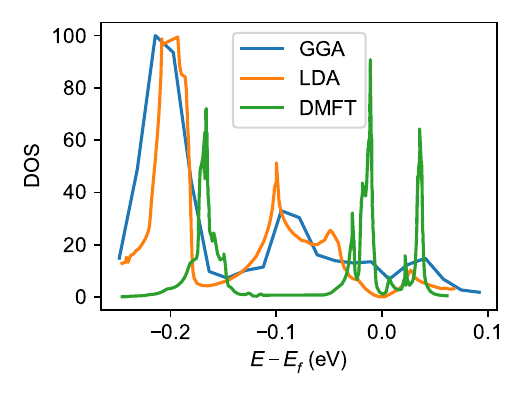}
	\caption{
            Comparison between the DOS of the band structure calculated with GGA in the present work (blue), and that computed with the Local Density Approximation (orange) and Dynamical Mean Field Theory (green) in Ref.~\onlinecite{Thunstroem2021}.
	}	
	\label{fig:dos_comparison}
\end{figure}

As shown in Fig.~\ref{fig:dos_comparison}, the DOS we computed with GGA resembles the data obtained with LDA in Ref.~\onlinecite{Thunstroem2021}.
Furthermore, the DMFT spectral function is reminiscent of the DOS obtained with DFT, and the main effect of electron interactions is the further flattening of heavy bands around the Fermi level.
This results corroborates the fact that the electronic structure of \smbsix{} can be described in terms of strongly-renormalized heavy and dispersive bands, and encourages us to restrict our numerical analysis to DFT level.

\section{Excluding potential band-crossings on high-symmetry lines, planes and generic points} \label{app: crossings in SmB6}

In this appendix, we discuss the potential existence of accidental band crossings between valence and conduction bands in \smbsix.
For that, we first need to show how symmetries constrain the form of the Hamiltonian on a \bk-point in the BZ.

Let $\hat{H}(\bs{k})$ be the Hamiltonian operator restricted to bands that potentially cross at particular \bk-point. We consider that the states on these bands transform according to  the same physically irreducible representation (pirrep) $D^{\bs{k}}$, as otherwise they could not hybridize. $\hat{H}(\bs{k})$ can be written in the following form
\begin{equation} \label{eq: general H(k) operator}
	\hat{H}(\bs{k}) 
		= \sum_{ij} h_{ij}(\bs{k}) \op{i}{j},
\end{equation}
where $\ket{i}$ and $\ket{j}$ run over the states adapted to the symmetry of the pirrep of the bands that (potentially) cross. 
For a symmetry operation $g_{\bs{k}}$ in the little group of \bk \ the hamiltonian must satisfy 
\begin{equation} \label{eq: H (k) commutation}
\begin{split}
g_{\bs{k}}^{-1} \hat H (\bs{k}) g_{\bs{k}}
& = \sum_{ij} h_{ij} (\bs{k}) g_{\bs{k}}^{-1} \op{i}{j} g_{\bs{k}} \\
& = \sum_{ijmn} h_{ij} (\bs{k}) 
    [D^{\bs{k}\dagger}(g_{\bs{k}})]_{mi}  
    \op{m}{n} 
    D^{\bs{k}}_{jn} (g_{\bs{k}}) \\
\end{split}
\end{equation}
Since $\hat H (\bs{k})$ must commute with all the operations in the little group of \bk, the matrix elements $h_{ij}(\bs{k})$ must satisfy the following equation
\begin{equation} \label{eq: relations h_ij}
	h_{mn}(\bs{k}) 
	= \sum_{ij} [D^{\bs{k}\dagger}(g_{\bs{k}})]_{mi}
	  h_{ij} (\bs{k})   
	  D^{\bs{k}}_{jn} (g_{\bs{k}}) 
\end{equation}
Eq.~\eqref{eq: relations h_ij} might set constrains on the matrix elements $h_{ij} (\bs{k})$ and therefore describes how crystal symmetries determine the form of $\hat H (\bs{k})$.

Let $\theta$ be an antiunitary symmetry in the little group of \bk.
The operator in the pirrep $D^{\bs{k}}$ of such a symmetry can be written as the combination of a unitary matrix $U$ and the complex conjugation operator $\mathcal{K}$, i.e. $D^{\bs{k}}(\theta) = U\mathcal K$.
$\hat H (\bs{k})$ should also commute with $\theta$, which imposes the following constrain on the matrix elements $h_{ij} (\bs{k})$

\begin{equation} \label{eq: relations antiunitary h_ij}
	h_{mn}^{*}(\bs{k}) 
	= \sum_{ij} U^{\dagger}_{mi}
	h_{ij} (\bs{k})   
	U_{jn} 
\end{equation}

In the remaining of this appendix we will focus on the line $\Delta(\Gamma X)$ whose points have coordinates $(0,k_y,0)$.
A detailed analysis for the rest of symmetry lines, planes  and generic points can be found in Ref.~\onlinecite{niretesia}.
We will begin deriving the most general form of $\hat H (\bs{k})$ that is compatible with the symmetry of the little group of \bk-points in this line, based on Eqs.~\eqref{eq: relations h_ij} and \eqref{eq: relations antiunitary h_ij}.

\subsubsection*{Line $\Delta: (0,k_y,0)$}

The closing of the gap between valence and conduction bands would involve a touching of states that transform as the pirreps $\bar{\Delta}_6$.
We denote $\ket{i}$ and $\ket{j}$ the symmetry-adapted states of one of the $\Delta_6$ pirreps and $\ket{i'}$ and $\ket{j'}$ those of the other $\Delta_6$ pirrep.

Let us consider the matrix elements $h_{ii}(\Delta)$, $h_{ii'}(\Delta)$ and $h_{ij}(\Delta)$ in Eq.~\eqref{eq: general H(k) operator}. 
Instead of checking case by case the constrains set by all symmetries in $G_\Delta$, it is sufficient to consider only the action of the generators (see Tab.~\ref{tab: line delta}).
In particular, applying Eq.~\eqref{eq: relations h_ij} for the four-fold rotation $C_{4y}$ yields
\begin{equation}
h_{ij} =  -i h_{ij} \Rightarrow h_{ij} = 0.
\end{equation}
and, similarly,  $h_{i'j'} = h_{ij'} = h_{i'j} = 0$. The action of the reflection $m_{z}$ yields
\begin{equation}
\begin{split}
& h_{ii} = h_{jj}, \\	
& h_{ii'} = h_{jj'}.
\end{split}
\end{equation}
together with  $h_{i'i'} = h_{j'j'}$ and $h_{i'i} = h_{j'j}$.  

As  points on the  $\Delta$ line are not time-reversal invariant, we cannot choose this operation as the antiunitary representative. 
Nevertheless, we can select the combination of inversion and time-reversal symmetry, i.e. $I \mathcal{T}$, which does belong to the little group. 
By applying the action of the unitary part $U$ of $I \mathcal{T}$ (see Tab.~\ref{tab: line delta}) in Eq.~\eqref{eq: relations antiunitary h_ij}, we obtain the constrain that all matrix elements of $\hat{H} (\Delta)$ must be real functions. 
Altogether, the most general form of the matrix $H(\Delta)$ compatible with the symmetries in the basis $\{ \ket{i}, \ket{i'}, \ket{j}, \ket{j'} \}$ is the following

\begin{equation} \label{eq: H(Delta)}
H(\Delta) = 
\begin{pmatrix}
		a(k_y) & b(k_y) & 0 & 0 \\
		b(k_y) & a'(k_y) & 0 & 0 \\
		0 & 0 & a(k_y) & b(k_y) \\
		0 & 0 & b(k_y) & a'(k_y)
\end{pmatrix},
\end{equation}
where $a(k_y)$, $a'(k_y)$ and $b(k_y)$ are real functions whose particular form depends on the microscopic details of the crystal. The eigenvalues of this matrix are 

\begin{equation} \label{eq: eigenvalues H(Delta)}
	E_{\pm} (k) = \dfrac{1}{2} [a(k_y) + a'(k_y)] \pm \sqrt{ \dfrac{1}{2}[a(k_y) - a'(k_y)]^2 + b^2(k_y) }.
\end{equation}

The square root in Eq.~\eqref{eq: eigenvalues H(Delta)} should vanish to have a band crossing. 
This requires $a(k_y) = a'(k_y)$ and $b(k_y)=0$ to be satisfied simultaneously.
The first condition is met at the intersection $k_y$ of two curves, whereas the second equation defines the point $k_{y}'$.
The coincidence $k_y = k_{y}'$ requires fine-tuning of the material's microscopic features, thus it is impossible that two bands that transform as $\bar \Delta_6$ cross.

Since the Hamiltonian corresponding to $\bar \Delta_7$ bands is identical to \EQ{eq: H(Delta)}, it is impossible to have a crossing between $\bar \Delta_7$ bands without the infinitely-accurate tunning of the system's microscopic parameters, which would be unrealistic.

\begin{table}[t] 
	\caption{Matrices for the generators of the little cogroup of $G_\Delta$ in the representations $\bar{\Delta}_6$ and $\bar{\Delta}_7$.}
	\label{tab: line delta}
	\begin{tabular}{l c c c}
		\hline \hline
		\textrm{pirrep} \rule{2ex}{0pt}
		& $\{ 4_{010}^+ | 000 \}$ 
		& $\{ m_{001} | 000 \}$ 
		& $\mathcal{T}\{ I | 000 \}$ 
		\\ \hline \noalign{\vskip 2mm} 
		$\bar\Delta_6$ 
		& $\twobytwo{e^{i 3\pi/4}}{0}{0}{e^{-i 3\pi/4}}$ 
		& $\twobytwo{0}{e^{i \pi/4}}{e^{i 3\pi/4}}{0}$ 
		& $\twobytwo{0}{-1}{1}{0}$
		\\ 
		\noalign{\vskip 2mm}
		$\bar\Delta_7$ 
		& $\twobytwo{e^{-i \pi/4}}{0}{0}{e^{i\pi/4}}$ 
		& $\twobytwo{0}{e^{-i 3\pi/4}}{e^{-i \pi/4}}{0}$ 
		& $\twobytwo{0}{-1}{1}{0}$ \\
		\noalign{\vskip 2mm} \hline \hline
	\end{tabular}
\end{table}

\section{Construction of the tight-binding model} \label{app: construction TB}

Let us denote the tight-binding basis states $\ket{p_{\bs{R},i,\sigma}}$, where $\bs{R}$ is the lattice vector of the unit cell, $p=g,u$ stands for the parity of the corresponding irrep, $i=1,2$ labels the state within the basis of the irrep $E_p$ and $\sigma$ is the spin-degree of freedom--for example, $\ket{g_{\bs{R},2,\uparrow}}$ is the basis state of $\bar F_g$ in the cell $\bs{R}$ constructed as the product of the second basis state of $E_g$ and the $\uparrow$-spin state.
The transformation of these states under a symmetry $h \in m\bar{3}m$ is described by the following expression

\begin{equation} \label{eq: transformation TB states RS}
	h \ket{p_{\bs{R},i,\sigma}} = [E_p(h)]_{i'i} S_{\sigma'\sigma}(h)  \ket{p_{(h\bs{R}),j,\sigma'}},
\end{equation}
where $E_p(g)$ is the matrix of $h$ in the representation $E_p$.
We will use greek letters to denote the degrees of freedom corresponding to the irreps, except for the parity.
The matrix of $h$ in Eq.~\eqref{eq: transformation TB states RS} will be written accordingly as $V_{\alpha'\alpha} = [E_p(h)]_{i'i} S_{\sigma'\sigma}(h)$, with $\alpha=(i\sigma)$ and $\alpha'=(i'\sigma')$.
Then, the matrix elements of the Hamiltonian in the basis of tight-binding states defined in real space can be written as:

\begin{equation}
	H_{p\alpha,p'\alpha'}(\bs{R}) = \mel{p_{\bs{R},\alpha}}{H}{p'_{\bs{0},\alpha'}},
\end{equation}
Here, we only consider amplitudes for hoppings from the unit cell at the origin. The remaining amplitudes can be related to these through translations by vectors of the lattice. The fact that the Hamiltonian must be invariant under all space-group symmetries, together with Eq.~\eqref{eq: transformation TB states RS}, leads to the following relation between hopping amplitudes

\begin{equation} \label{eq: transformation matrix elements RS}
	\begin{split}
		H_{p\beta, p'\beta'}(h\bs{R})
		= 
		V_{\beta\alpha}(h) H_{p\alpha, p'\alpha'}(\bs{R}) V^\dagger_{\alpha'\beta'}(h).
	\end{split}
\end{equation}
For certain symmetry operations, this relation could further set constrains on some matrix elements, reducing the number of independent parameters needed to describe the considered couplings.

\begin{figure} 
	\centering
	\includegraphics[width=1.0\linewidth]{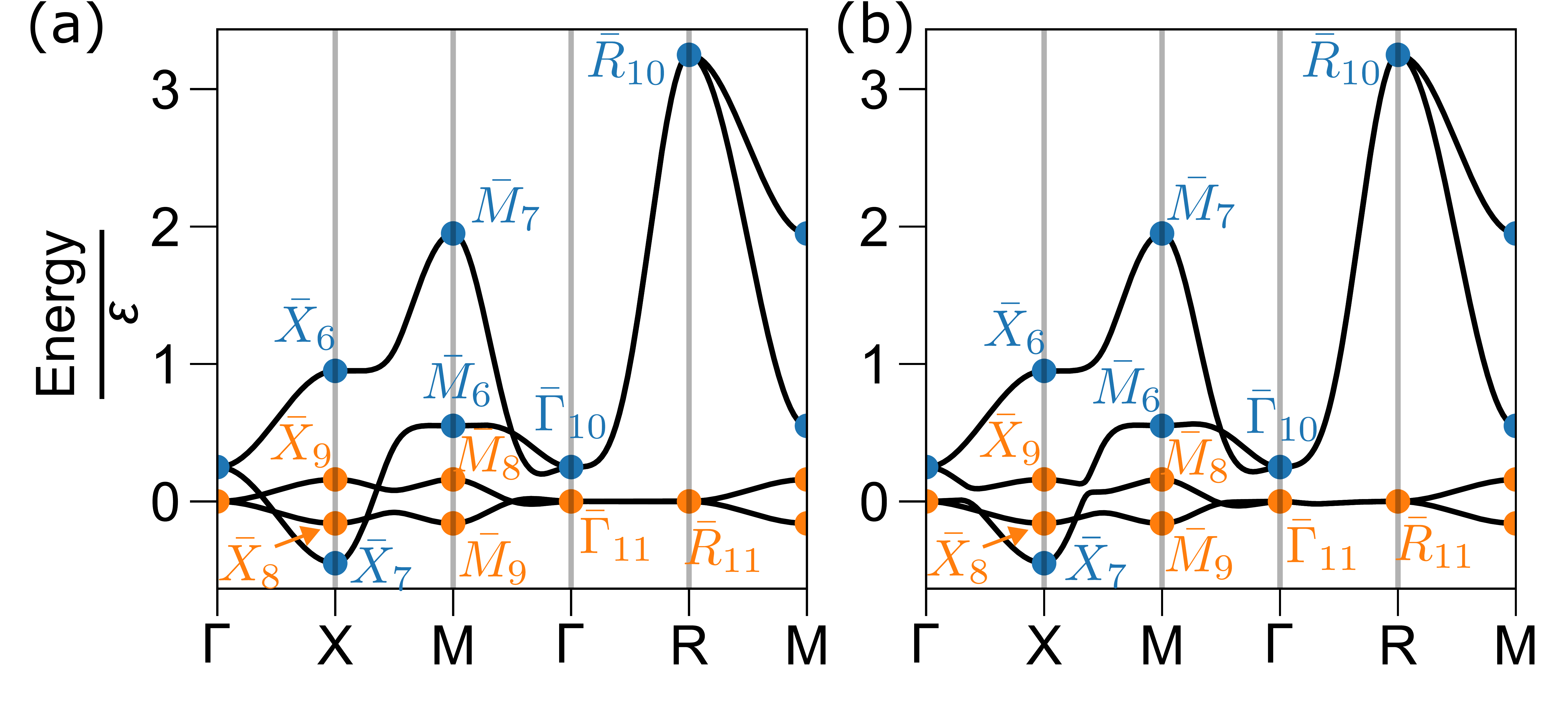}
	\caption{Tight-binding band structure of \smbsix{} with $t_{1}^{NN}/\epsilon_d = -1/2$, $t_2/\epsilon_d = 1/8$, $t_3/\epsilon_d = 0.7/4$ and $t_4/\epsilon_d = -0.04$. The rest of hopping parameters are chosen to be zero. (a) Bands without hybridization between $f$ and $d$-orbitals. (b) Bands with hybridization $V=0.06$. Valence bands have in correspondence identical values for symmetry-indicators of topology as DFT valence bands calculations. 
        }
        \label{fig:SmB6 TB bands}
\end{figure}

Fig.~\ref{fig:SmB6 TB bands} shows the band structure of the tight-binding model, with values for the hopping parameters chosen as to reproduce the ordering of irreps in Fig.~\ref{fig:bandstructure}(b).
When the hybridization between $\bar{F}_g$ and $\bar F_u$ states is considered ($v>0$), the first four bands separate from the rest by a gap.
The set of irreps of these bands has in correspondence the same values for the symmetry-indicators of topology as the \textit{ab initio} valence bands.
In particular, they are characterized for having $z_{4\pi m} = 3$ and $z_8 = 5$.
Thus the tight-binding model presented here is able to reproduce the topological phase obtained via \textit{ab initio} calculations.

\section{Topological gaps between valence bands in \smbsix} \label{app: valence topological gaps}

In the main text we have focused on the multiple topological gaps that might yield close to the Fermi level the interplay between $d$ and $f$-bands.
Nevertheless, the origin of topological gaps is not restricted to these bands.
As it is shown in Tab.~\ref{tab:gaps}, the gaps between the $27^{\textrm{th}}$ and $28^{\textrm{th}}$ bands, as well as the $29^{\textrm{th}}$ and $30^{\textrm{th}}$ bands, are also topological.
These gaps involve B bands, and are of the order of 1 meV, or even smaller.
However, their topology does not contribute to the cumulative topology of the whole set of valence bands, due to the fact that the set of bands coming from B states is completely occupied and has an atomic limit.
Moreover, they are gaps located well below the Fermi level, thus accessing them experimentally might be more complicated.

\begin{table}[t] 
	\caption{Irreps of the isolated sets of valence bands in the MBJ approximation and  their cummulative topology. The first column contains the index of the last band in each set, with the energy increasing with the band number. The last column gives  the values for the symmetry indicators of the cummulative topology.}
	\label{tab:gaps}
	\begin{tabular}{c c c c c c}
		\hline \hline
                band & $\Gamma$ & X & M & R & $(z_{2w,1}, z_{2w,2}, z_{2w,3}, z_4, z_2, z_8, z_{4\pi m} )$ \\ 
                \hline
                2 & $\bar\Gamma_6$ & $\bar X_6$ & $\bar M_7$ & $\bar R_6$ & $(0,0,0,0,0,0,0)$ \\ \hline
                4 & $\bar\Gamma_8$ & $\bar X_8$ & $\bar M_9$ & $\bar R_8$ & $(0,0,0,0,0,0,0)$ \\ \hline
                8 & $\bar\Gamma_{11}$ & $\bar X_8$ & $\bar M_8$ & $\bar R_{11}$ & $(0,0,0,0,0,0,0)$ \\ 
                 & & $\bar X_9$ & $\bar M_9$ & &  \\ \hline
                 22 & $\bar\Gamma_6$ & $\bar X_8$ & $\bar M_6$ & $\bar R_9$ & $(0,0,0,0,0,0,2)$\\
                & $\bar\Gamma_{6}$ & $\bar X_6$ & $\bar M_9$ & $\bar R_{10}$ &  \\ 
                   & $\bar\Gamma_{8}$ & $\bar X_6$ & $\bar M_8$ & $\bar R_{7}$ &  \\ 
                   & $\bar\Gamma_{11}$ & $\bar X_7$ & $\bar M_6$ & $\bar R_{8}$ &  \\ 
                   & $\bar\Gamma_{10}$ & $\bar X_8$ & $\bar M_8$ & $\bar R_{11}$ &  \\ 
                   &                   & $\bar X_9$ & $\bar M_7$ &  &  \\ 
                   &                   & $\bar X_8$ & $\bar M_7$ &  &  \\ \hline
                28 & $\bar\Gamma_{10}$ & $\bar X_9$ & $\bar M_6$ & $\bar R_{10}$ & $(1,1,1,1,1,5,3)$ \\ 
                   & $\bar\Gamma_{7}$ & $\bar X_6$ & $\bar M_9$ &$\bar R_{7}$  &  \\ 
                 &  & $\bar X_7$ & $\bar M_8$ &  & \\ \hline
                30 & $\bar\Gamma_{9}$ & $\bar X_9$ & $\bar M_8$ & $\bar R_{9}$ & $(1,1,1,1,1,5,3)$ \\ \hline
                34 & $\bar\Gamma_{11}$ & $\bar X_9$ & $\bar M_9$ & $\bar R_{11}$ & $(1,1,1,1,1,5,3)$  \\ 
                   & & $\bar X_8$ & $\bar M_8$ &  &  \\ 
                \hline \hline
	\end{tabular}
\end{table}

\bibliography{biblio_kondo}

\begin{thebibliography}{75}
\bibitem [{not()}]{note_hamiltonian}%
  \BibitemOpen
  \href@noop {} {}\bibinfo {note} {The Hamiltonian considered here is not the
  most general symmetry-consistent Hamiltonian that could be constructed with
  couplings between these states. It is indeed a simplification obtained by
  considering that $f-f$ hoppings to next-nearest neighbors are governed by a
  real parameter.}\BibitemShut {Stop}%
\end{thebibliography}
\end{document}